# Autonomous Orbit Determination Using Epoch-Differenced Gravity Gradients and Starlight Refraction

Pei Chen, Tengda Sun, Xiucong Sun[*]

*School of Astronautics, Beihang University, Beijing 100191, China*

**Abstract**

Autonomous orbit determination via integration of epoch-differenced gravity gradients and starlight refraction is proposed in this paper for low-Earth-orbiting satellites operating in GPS-denied environments. The starlight refraction can compensate for the significant along-track position error using solely gravity gradients and benefit from the integration in view of accuracy improvement in radial and cross-track position estimates. The between-epoch differencing of gravity gradients is employed to eliminate slowly varying measurement biases and noises near the orbit revolution frequency. The refraction angle is directly used as measurement and its Jacobian matrix is derived from an implicit observation equation. An information fusion filter based on sequential extended Kalman filter is developed for the orbit determination. Truth-model simulations are used to test the performance of the algorithm and the effects of differencing intervals and orbital heights are analyzed. A semi-simulation study using actual gravity gradient data from the Gravity field and steady-state Ocean Circulation Explorer (GOCE) combined with simulated starlight refraction measurements is further conducted and a three-dimensional position accuracy of better than 100 m is achieved.

## 1. Introduction

Knowledge of position and velocity is essential to satellite operation, such as command and control, preliminary instrument calibration, as well as mission planning. Traditional orbit determination strongly relies on ground-based infrastructure and is not suitable for future autonomous space missions. Recent years have witnessed a series of studies on autonomous orbit determination for low-Earth-orbiting (LEO) satellites. Nagarajan et al.[1] investigated the use of low-cost Earth scanners combined with known attitude information for orbit estimation. The development of Global Positioning System (GPS) has contributed to the progress of satellite navigation using onboard GPS receivers.[2,3] Novel astronomical methods (not limited to LEOs) which utilize the observations of x-ray pulsars, γ-ray photons, starlight refraction, as well as the Earth's magnetic field have also been proposed and studied.[4-7] The realization of autonomous orbit determination can reduce the burden of ground stations and free operators to handle more pressing problems. In addition, the survival ability of spacecraft can also be enhanced.

---

[*] Corresponding author. Tel.: +86 10 82316535.
  E-mail address: sunxiucong@gmail.com (X. Sun).

More recently, Chen et al.[8] proposed to use the observation of the Earth's gravity gradients for spacecraft navigation. As the second-order gradient of the gravitational potential, the gravity gradient tensor (GGT) varies with position and orientation relative to the Earth reference frame. When high-precision attitude information is provided, the position or orbit trajectory can be obtained by matching the observations with an existing gravity model. An eigendecomposition method was presented to translate GGT into position using the $J_2$ gravity model. Sun et al.[9] further considered the effects of gravity gradient biases in actual measurements and developed an adaptive hybrid least-squares batch filter to simultaneously estimate the orbital states and unknown biases. Application to the European Space Agency (ESA)'s Gravity field and steady-state Ocean Circulation Explorer (GOCE) indicates the practicality of the method. A different orbit determination strategy based on the extended Kalman filter (EKF) was also established in Ref.[10] and a comparable accuracy was demonstrated for GOCE. The gravity gradient based orbit determination is immune to signal blockage and spoofing encountered in GPS navigation.[11] In contrast to the Earth's magnetic field, the Earth's gravitational field is not affected by solar activities and only endures secular variations due to the its interior changes.[12,13] The typical position errors of orbit determination using magnetometer measurements range from a few to tens of kilometers,[7,14,15] whereas an orbital position accuracy of hundreds of meters has been achieved with GOCE gravity gradient measurements.

The case study of GOCE orbit determination conducted in Ref.[9] revealed a phenomenon that the along-track position component endured much larger error than the radial and cross-track components. Specifically, the radial and cross-track position errors were 10.4 m and 22.8 m, respectively, whereas the along-track position error was over one order of magnitude larger (677.0 m) and restricted the overall orbit determination accuracy. This non-uniform error distribution was attributed to the poor observability of the bias on the $V_{xz}$ gravity gradient component and thus can be considered as an inherent characteristic of the system. An initial calibration based on ground-based tracking could be used to improve the along-track position accuracy. However, the measurement biases are drifting slowly and any minor estimation error of the drift rate would lead to accumulated position error over a long time. Therefore, frequent calibration should be carried out in practice. An alternative compensation method to ground-based calibration is integrated navigation with a second sensor. Among the other autonomous orbit determination techniques, the starlight refraction is an ideal choice. First, the starlight refraction as an astronomical method, similar to gravity gradiometry, is non-emanating and nonjammable and can also be used in GPS-denied environments. Second, recent studies on starlight refraction based navigation have shown that position errors of one or two hundred meters could be achieved with a refraction angle measurement accuracy of 1 arcsec.[6,16,17] This position accuracy is comparable with that of the gravity gradient based orbit determination and in particular is higher for the along-track component and lower for radial and cross-track components. Thus a win-win mechanism could be set up. Last but not least, the star sensor as the main measurement unit for starlight refraction is also required in gravity gradient based orbit determination to provide high-precision attitude information. Instrument integration can be relatively easily accomplished by installation of an additional refraction star sensor.

An early concept of navigation using starlight refraction was presented in Ref.[18]. Following researchers from the Charles Stark Draper Laboratory conducted error analysis studies of autonomous navigation based on EKF and concluded that a position error of less than 100 m would be possible.[19,20] With the improvement of starlight atmospheric refraction model accuracy as well as the precision of star sensors, new contributions to starlight refraction based navigation have been made in these years. Wang et al.[21] established an empirical model of atmospheric refraction for a continuous range of height ranging from 20 km to 50 km in terms of atmospheric temperature, pressure, density, as well as density scale height. A good consistency is found between this empirical model and the actual observed data. The same empirical model was adopted in Ref.[17]. Instead of using refraction apparent height as measurement, the refraction angle was directly used in order to eliminate the effects of nonlinear error propagation. The unscented Kalman filter (UKF) was utilized to deal with the nonlinearity of the measurement equation and a position accuracy of better than 100 m has been achieved for a LEO satellite at an altitude of 786 km with a detectable stellar magnitude of 6.95.

The present study investigates the possible integration of gravity gradiometry and starlight refraction for autonomous orbit determination of LEO satellites. As mentioned earlier, the integration will not only compensate for the large along-track position error encountered in orbit determination using only gravity gradients but also increase the radial and cross-track position accuracies for the starlight refraction based navigation. In addition, the effects of starlight refraction data outages due to invisibility of refracted stars during some periods can also be reduced via the integration. Different with the estimation method presented in Ref.[9], a sequential filter rather than a batch filter is utilized in order to satisfy the real-time or near real-time requirements for autonomous orbit determination. The gravity gradients are differenced using measurements from neighbor epochs to eliminate the slowly varying biases and noises near the orbit frequency. Compared to the augmented state filter given in Ref.[10], the dimensionality of the state vector is reduced. As for starlight refraction, the refraction angle is directly used as measurement and its Jacobian matrix is derived in this study. An information fusion filter which sequentially processes epoch-differenced gravity gradient (EDGG) and starlight refraction angle (SRA) measurements via EKF has been developed for orbit determination. The algorithm is applied to both simulated data and the actual gravity gradient data from GOCE. The performance of the integration navigation is evaluated and compared with those of the methods using solely EDGG or SRA measurements.

The structure of this paper is organized as follows. Section 2 briefly reviews the basic principles of gravity gradiometry and starlight refraction. Section 3 presents the orbital dynamic model, the measurement models of EDGG and SRA observations, as well as the information fusion filter design. Simulation results and several important factors are presented and analyzed in Section 4. The semi-simulation based on actual GOCE gravity gradiometry data and simulated SRA measurements is presented in Section 5. Conclusions of this study are drawn in Section 6.

## 2. Brief review of gravity gradiometry and starlight refraction

*2.1. Gravity gradiometry*

The gravity gradient tensor $\boldsymbol{\Gamma}$ consists of the second-order partial derivatives of the gravitational potential $U$ with respect to the position vector. Its coefficient matrix with respect to a specific coordinate reference frame takes the following form

$$\{\boldsymbol{\Gamma}\}_a = \{\nabla_r^2 U\}_a = \begin{bmatrix} \Gamma_{xx}^{(a)} & \Gamma_{xy}^{(a)} & \Gamma_{xz}^{(a)} \\ \Gamma_{yx}^{(a)} & \Gamma_{yy}^{(a)} & \Gamma_{yz}^{(a)} \\ \Gamma_{zx}^{(a)} & \Gamma_{zy}^{(a)} & \Gamma_{zz}^{(a)} \end{bmatrix} \quad (1)$$

where $\Gamma_{\alpha\beta} = \frac{\partial^2 U}{\partial \alpha \partial \beta}$ $\forall \{\alpha, \beta\} \in \{x, y, z\}$ and $x$, $y$, and $z$ are the components of the position vector $r$ expressed in frame $a$. The unit of gravity gradients is Eötvös, denoted by the symbol E. In the international system of units (SI), 1 E = $10^{-9}$ s$^{-2}$. The continuity of the gravitational potential ensures that $\{\boldsymbol{\Gamma}\}_a$ is symmetric and the Laplace's equation restricts its trace to be zero. Thus, there are only five independent terms in $\{\boldsymbol{\Gamma}\}_a$.

The coefficient matrix of GGT depends on the choice of reference system. The relationship between GGTs expressed in two different frames is given as follows

$$\{\boldsymbol{\Gamma}\}_b = \boldsymbol{C}_a^b \{\boldsymbol{\Gamma}\}_a (\boldsymbol{C}_a^b)^T \qquad (2)$$

where the symbol $b$ denotes a second frame and $\boldsymbol{C}_a^b$ is the coordinate rotation matrix from frame $a$ to $b$. More details about the characteristics of gravity gradients can be found in Heiskanen and Moritz.[22]

The GGT can be measured by a gravity gradiometer which usually comprises three orthogonal pairs of high-precision three-axis accelerometers, as shown in Fig. 1. The output of each accelerometer is given by

$$\boldsymbol{a}_i = \left(\boldsymbol{\Omega}^2 + \dot{\boldsymbol{\Omega}} - \boldsymbol{\Gamma}\right)\boldsymbol{r}_i + \boldsymbol{d}, \ i = 1, 2, ..., 6 \qquad (3)$$

where $i$ is the identifier of the accelerometer, $\boldsymbol{r}_i$ is the vector from the gradiometer center to the position of the $i$th accelerometer, $\boldsymbol{\Omega}$ is the cross product matrix of the angular velocity of the gradiometer, $\boldsymbol{\Omega}^2$ and $\dot{\boldsymbol{\Omega}}$ represent the square and time derivative of $\boldsymbol{\Omega}$, and $\boldsymbol{d}$ is the non-gravitational acceleration at the gradiometer center attributed to atmospheric drag, solar radiation pressure and thruster forces.

The differences of accelerometer outputs can remove the common-mode non-gravitational accelerations

$$\boldsymbol{a}_{d,ij} = \frac{1}{2}(\boldsymbol{a}_i - \boldsymbol{a}_j) = \frac{1}{2}\left(\boldsymbol{\Omega}^2 + \dot{\boldsymbol{\Omega}} - \boldsymbol{\Gamma}\right)\boldsymbol{L}_{ij} \qquad (4)$$

where $ij \in \{14, 25, 36\}$ represents the index of the accelerometer pairs and $\boldsymbol{L}_{ij}$ is the vector from the $j$th to the $i$th accelerometer. Combine the three differenced accelerations to form a matrix equation

$$\begin{aligned}\boldsymbol{A} &= \begin{bmatrix} \boldsymbol{a}_{d,14} & \boldsymbol{a}_{d,25} & \boldsymbol{a}_{d,36} \end{bmatrix} \\ &= \frac{1}{2}\left(\boldsymbol{\Omega}^2 + \dot{\boldsymbol{\Omega}} - \boldsymbol{\Gamma}\right)\boldsymbol{L}\end{aligned} \qquad (5)$$

where $\boldsymbol{L} = \begin{bmatrix} \boldsymbol{L}_{14} & \boldsymbol{L}_{25} & \boldsymbol{L}_{36} \end{bmatrix}$. Based on the symmetry of $\boldsymbol{\Omega}^2$ and $\boldsymbol{\Gamma}$ and the skew-symmetry of $\dot{\boldsymbol{\Omega}}$, the GGT can be retrieved by

$$\boldsymbol{\Gamma} = -\left[\left(\boldsymbol{A}+\boldsymbol{A}^T\right)\boldsymbol{L}^{-1} - \boldsymbol{\Omega}^2\right] \tag{6}$$

It should be noted that the numerical readings of the gradiometer correspond to the coefficient matrix of GGT in the gradiometer reference frame (GRF), which is defined and materialized by the three orthogonal baselines of accelerometers.

Taking the accelerometers' intrinsic biases and noises, the gradiometer geometric imperfections as well as the angular velocity estimation errors into account, the actual observed GGT is given by

$$\boldsymbol{V} = \left\{\boldsymbol{\Gamma}\right\}_g + \boldsymbol{B} + \boldsymbol{N}_{Orb} + \boldsymbol{N}_w \tag{7}$$

where the symbol 'g' denotes the GRF frame, $\boldsymbol{V}$ is the numerical reading of the gradiometer in matrix form, $\boldsymbol{B}$ is a slowly drifting bias matrix, $\boldsymbol{N}_{Orb}$ is the matrix containing noises near the orbit frequency, and $\boldsymbol{N}_w$ is the matrix containing white noises. The detailed analysis of sources of GGT measurement error is given in Ref.[9].

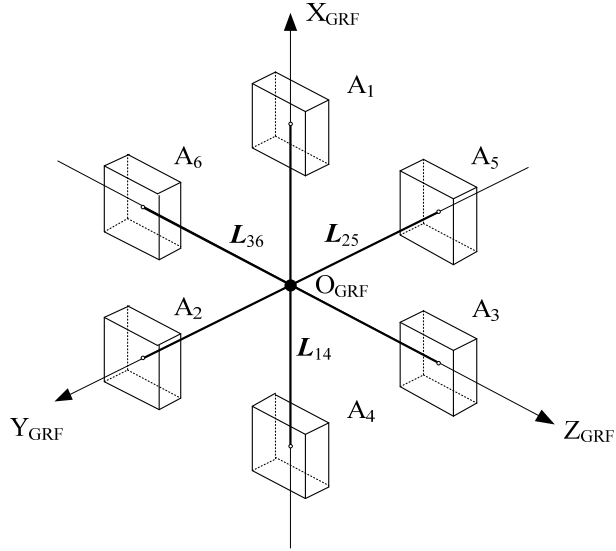

**Fig. 1.** Schematic illustration of configuration of the gradiometer and arrangement of the six accelerometers.

*2.2. Starlight refraction*

As depicted in Fig. 2, the passage of starlight through the Earth's atmosphere bends the rays inward due to atmospheric refraction, which causes a higher apparent position of the star than its true position viewed from a LEO spacecraft. The refraction effect is greatest near the Earth's surface and decreases in an exponential manner with increasing height.[19] Let $h_g$ denote the actual refraction tangent height of the star. According to Ref.[18], the refraction angle $R$ can be approximately given by

$$R = k(\lambda)\rho_g \sqrt{\frac{2\pi(R_E + h_g)}{H_g}} \tag{8}$$

where $k(\lambda)$ is the dispersion parameter and only a function of the wavelength of the starlight $\lambda$, $\rho_g$ is the atmospheric density at height $h_g$, $R_E$ is the reference equatorial radius of the Earth, and $H_g$ is the density scale height.

According to Ref.[20], the relationship between the apparent height $h_a$ and the refraction tangent height $h_g$ is

$$h_a = h_g + k(\lambda)\rho_g R_E \tag{9}$$

Thus the relationship between $h_a$ and $R$ can be obtained by combination of Eqs. (8) and (9). An empirical model has been given in Ref.[17] to express their functional relationship

$$h_a = -21.74089877 - 6.441326 \ln R + 69.21177057 R^{0.9805} \tag{10}$$

where the units of $R$ and $h_a$ are radian and kilometer, respectively.

The geometric relationship between the apparent height and the satellite position is

$$h_a \approx \sqrt{r^2 - u^2} + u \tan R - R_E \tag{11}$$

where $r = \|\mathbf{r}\|$, $\mathbf{r}$ is the satellite position vector, $u = |\mathbf{r} \cdot \mathbf{u}_s|$, $\mathbf{u}_s$ is the unit vector of the star before refraction. The implicit relationship between the refraction angle and the satellite position can thus be obtained from Eqs. (10) and (11) as follows

$$\sqrt{r^2 - u^2} + u \tan R - R_E = -21.74089877 - 6.441326 \ln R + 69.21177057 R^{0.9805} \tag{12}$$

According to Wang et al.[21], the starlight refraction angle at tangent height of 25 km calculated using the above empirical model has an error of 0.2 arcsec compared with observed data.

The starlight refraction angle can be measured by employing two onboard star sensors.[17] The first star sensor is zenith-pointing and is used to observe non-refracted stars, called the attitude star sensor. The second one has its optical axis pointing to the Earth's limb and is used to observe refracted stars, called the refraction star sensor. The attitude information deduced from the first star sensor can be used to generate a simulated star image for the second one. By comparing the simulated and actual star images, the refraction angle can be directly obtained.

Let $(x_a, y_a)$ and $(x_b, y_b)$ denote the coordinates of one refracted star in the simulated and actual star images, respectively. The refraction angle is

$$R = \arccos\left[\frac{x_a x_b + y_a y_b + f^2}{\sqrt{(x_a^2 + y_a^2 + f^2)(x_b^2 + y_b^2 + f^2)}}\right] \quad (13)$$

where $f$ is the focus of the second star sensor.

The number of refracted stars observed per orbit period is closely related to the installed angle of the refraction star sensor. This study employs the optimal installation strategy which has been proposed in Ref.[6] for observing refracted stars with tangent heights ranging from 20 km to 50 km. Let $\theta_{FOV}$ represent the field of view (FOV) of the refraction star sensor. The optimal installed angle is given by

$$\theta = \frac{1}{2}\left\{\arccos\left[\frac{\cos\alpha}{\cos(\theta_{FOV}/2)}\right] + \arccos\left[\frac{\cos\beta}{\cos(\theta_{FOV}/2)}\right]\right\} \quad (14)$$

with

$$\alpha = \arcsin\left[(20\,km + R_E)/r\right] \quad (15)$$

$$\beta = \arcsin\left[(50\,km + R_E)/r\right] \quad (16)$$

where $\alpha$ and $\beta$ correspond to the minimum and maximum tangent heights of refracted stars.

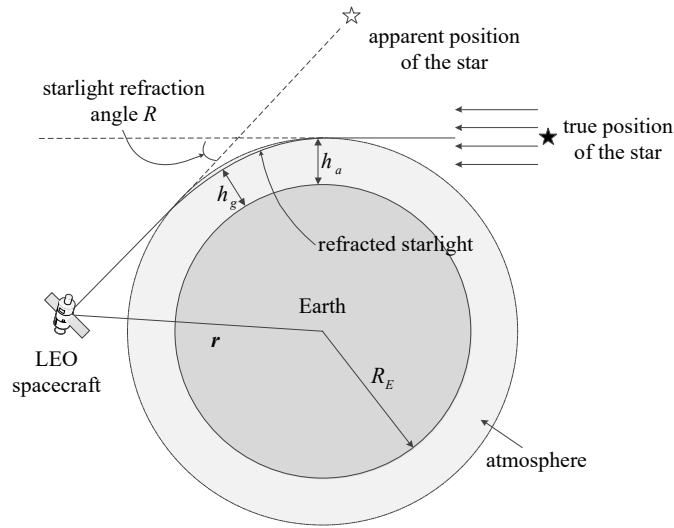

**Fig. 2.** Starlight refraction geometry for a LEO spacecraft.

# 3. Orbit determination algorithm

The orbit determination algorithm is used to estimate the position and velocity of the satellite from noisy observations which refer to epoch-differenced gravity gradients and starlight refraction angles in this paper. The algorithm is described by three parts: dynamic model of orbital motion, measurement models of EDGG and SRA, and the information fusion filter. The dynamic model is used to predict the satellite position and velocity as well as their covariances. The measurement models can be used to compute the modeled measurements and their partial derivatives with respect to the state variables. The information fusion filter combines the actual EDGG and SRA measurements and corrects the predicted states using the measurement innovations.

*3.1. Orbital dynamic model*

The orbital motion of a LEO satellite can be represented by the following differential equation expressed in the Earth-Centered Inertial (ECI) frame

$$\frac{d}{dt}\begin{bmatrix} \boldsymbol{r} \\ \boldsymbol{v} \end{bmatrix} = \begin{bmatrix} \boldsymbol{v} \\ \boldsymbol{f}(\boldsymbol{r},\boldsymbol{v},t) \end{bmatrix} + \begin{bmatrix} 0 \\ \boldsymbol{w}_t \end{bmatrix} \qquad (17)$$

where $\boldsymbol{r}$ and $\boldsymbol{v}$ are the position and velocity vectors, $\boldsymbol{f}(\cdot)$ is a 3-dimensional vector function representing the acceleration of deterministic forces, and $\boldsymbol{w}_t$ represents the remaining unmodeled perturbation acceleration and is assumed to be zero-mean white Gaussian noise.

In this study, the gravitational forces up to degree 20 and order 20 are modeled for the deterministic forces. The accelerations due to higher degree and order geopotential coefficients, the third-body gravitational attractions, and the non-gravitational forces are included into the noise $\boldsymbol{w}_t$. The standard deviation of $\boldsymbol{w}_t$ should reflect the actual accuracy of the dynamic model. Numerical analysis has been conducted to determine the standard deviation of $\boldsymbol{w}_t$ and shows that the values from $5 \times 10^{-4}$ m/s$^2$ to $5 \times 10^{-7}$ m/s$^2$ are appropriate for orbital heights ranging from 200 km to 2000 km.

The measurements are taken at discrete moments. The continuous dynamic model should be discretized before being used in the estimation algorithm. The discretized state model can be described as

$$\boldsymbol{x}_k = \boldsymbol{\varphi}(\boldsymbol{x}_{k-1}, t_k, t_{k-1}) + \boldsymbol{w}_{k-1} \qquad (18)$$

where $x_k$ and $x_{k-1}$ are the orbital states at $t_k$ and $t_{k-1}$. $\varphi(\cdot)$ is a 6-dimensional vector function and $w_k$ is the discrete process noise. The vector function $\varphi(\cdot)$ has no explicit expression and can only be numerically acquired with an ordinary differential equation solver. The state transition matrix $\Phi(t_k, t_{k-1})$ and the process noise covariance matrix $Q_k$ can be numerically obtained along with the integration of the orbital motion.

*3.2. Measurement models*

*3.2.1. Epoch-differenced gravity gradient*

The gravitational potential is usually modeled as a series of spherical harmonics[22]

$$U(r,\phi,\lambda) = \frac{GM}{r} \sum_{n=0}^{\infty} \left(\frac{R_E}{r}\right)^n \sum_{m=0}^{n} \overline{P}_{nm}(\sin\phi)\left[\overline{C}_{nm}\cos(m\lambda) + \overline{S}_{nm}\sin(m\lambda)\right] \quad (19)$$

where $r$, $\phi$, and $\lambda$ are the geocentric distance, latitude, and longitude of the position, $GM$ is the geocentric gravitational constant, $n$ and $m$ are the degree and order of the normalized spherical harmonic coefficients $\overline{C}_{nm}$ and $\overline{S}_{nm}$, and $\overline{P}_{nm}$ is the normalized associated Legendre function of the first kind. The International Earth Rotation and Reference Systems Service (IERS) recommends the Earth Gravitational Model 2008 (EGM2008) complete to degree 2190 and order 2159 as the conventional model and the International Terrestrial Reference Frame (ITRF) as the realization of the Earth-Centered Earth-Fixed (ECEF) frame for model representation.[23]

The gravity gradients in ECEF can be obtained by evaluating the second-order partial derivatives of $U$ with respect to position[9,10]

$$\{\boldsymbol{\Gamma}\}_e = \{\nabla_r^2 U\}_e \quad (20)$$

where the symbol '*e*' denotes the ECEF frame. In this study, a 120 × 120 subset of the EGM2008 gravity model is used to compute the modeled gravity gradient measurements. The accuracy of this truncated model is approximately 1 mE (milli-Eötvös) at height of 300 km.[10] The contributions of tidal effects are on the order of 0.1 mE and can be ignored.

As stated in Section 2.1, the gravity gradients are measured in the GRF frame. The relationship between $\{\boldsymbol{\Gamma}\}_e$ and $\{\boldsymbol{\Gamma}\}_g$ is

$$\{\boldsymbol{\Gamma}\}_g = \boldsymbol{C}_e^g \{\boldsymbol{\Gamma}\}_e (\boldsymbol{C}_e^g)^T \quad (21)$$

where $C_e^g$ is the coordinate rotation matrix from ECEF to GRF. The attitude star sensor provides accurate attitude information which can be used to derive the rotation matrix from ECI to GRF. The IERS provides accurate Earth orientation parameters which can be used to derive the rotation matric from ECEF to ECI. The rotation matrix from ECEF to GRF can be obtained by matrix multiplication.

Rewrite $\{\boldsymbol{\Gamma}\}_e$ and $\{\boldsymbol{\Gamma}\}_g$ into column vectors

$$\{\bar{\boldsymbol{\Gamma}}\}_i = \begin{bmatrix} \Gamma_{xx}^{(i)} & \Gamma_{yy}^{(i)} & \Gamma_{zz}^{(i)} & \Gamma_{xy}^{(i)} & \Gamma_{xz}^{(i)} & \Gamma_{yz}^{(i)} \end{bmatrix}^T, \quad i = e, g \tag{22}$$

The relationship between $\{\bar{\boldsymbol{\Gamma}}\}_e$ and $\{\bar{\boldsymbol{\Gamma}}\}_g$ can be written as

$$\{\bar{\boldsymbol{\Gamma}}\}_g = \boldsymbol{\Pi}_e^g \cdot \{\bar{\boldsymbol{\Gamma}}\}_e \tag{23}$$

where $\boldsymbol{\Pi}_e^g$ is a 6 × 6 projection matrix comprises elements from the coordinate rotation matrix $C_e^g$.[9]

By substituting Eq. (23) into the vector form of the gravity gradiometry equation, the gravity gradient measurement model can be obtained as follows

$$\boldsymbol{z}^{GG} = \boldsymbol{\Pi}_e^g \cdot \{\bar{\boldsymbol{\Gamma}}\}_e + \boldsymbol{b} + \boldsymbol{n}_{Orb} + \boldsymbol{n}_w \tag{24}$$

where $\{\bar{\boldsymbol{\Gamma}}\}_e$ contains the satellite position information, $\boldsymbol{b}$ is a bias vector corresponding to $\boldsymbol{B}$, and $\boldsymbol{n}_{Orb}$ and $\boldsymbol{n}_w$ are noise vectors corresponding to $\boldsymbol{N}_{Orb}$ and $\boldsymbol{N}_w$, respectively.

The epoch-differenced gravity gradient measurement model is obtained by differencing Eq. (24) between the current epoch and a previous epoch

$$\begin{aligned} \boldsymbol{z}_k^{EDGG} &= \boldsymbol{z}_k^{GG} - \boldsymbol{z}_{k-s}^{GG} \\ &= \left[\left(\boldsymbol{\Pi}_e^g \cdot \{\bar{\boldsymbol{\Gamma}}\}_e\right)\right]_k - \left[\left(\boldsymbol{\Pi}_e^g \cdot \{\bar{\boldsymbol{\Gamma}}\}_e\right)\right]_{k-s} + (\boldsymbol{b})_{k,k-s} + (\boldsymbol{n}_{Orb})_{k,k-s} + (\boldsymbol{n}_w)_{k,k-s} \\ &\approx \boldsymbol{h}^{EDGG}(\boldsymbol{x}_k, t_k) + (\boldsymbol{n}_w)_{k,k-s} \end{aligned} \tag{25}$$

where $s$ is the differencing interval, $(\cdot)_{k,k-s} = (\cdot)_k - (\cdot)_{k-s}$, and $\boldsymbol{h}(\cdot)$ represents the EDGG measurement function. The between-epoch differencing operation can remove the major part of measurement biases and noises near the orbit frequency. Thus, $(\boldsymbol{b})_{k,k-s} \approx \boldsymbol{0}$ and $(\boldsymbol{n}_{Orb})_{k,k-s} \approx \boldsymbol{0}$. Let $\boldsymbol{R}^{GG}$ denote the covariance matrix of the noise vector $\boldsymbol{n}_w$. The covariance matrix of $(\boldsymbol{n}_w)_{k,k-s}$ can be given as

$$\boldsymbol{R}^{EDGG} = 2\boldsymbol{R}^{GG} \tag{26}$$

$R^{GG}$ is a diagonal matrix since that the six gravity gradient components are measured independently by the gradiometer.

Let $T_e$ denote the partial derivative matrix of $\{\bar{\Gamma}\}_e$ with respect to position in ECEF.[9] The partial derivative matrix of $z^{GG}$ with respect to $r$ can be written as

$$H_r^{GG} = \Pi_e^g \cdot T_e \cdot C_i^e \tag{27}$$

where the symbol '$i$' here represents the ECI frame. The partial derivative matrix of $z_k^{EDGG}$ with respect to $r_k$ can be given by

$$H_{r,k}^{EDGG} = H_{r,k}^{GG} - H_{r,k-s}^{GG} \Phi_{rr}(t_{k-s}, t_k) \tag{28}$$

And the partial derivative matrix of $z_k^{EDGG}$ with respect to $v_k$ is

$$H_{v,k}^{EDGG} = -H_{r,k-s}^{GG} \Phi_{rv}(t_{k-s}, t_k) \tag{29}$$

where $\Phi_{rr}(t_{k-s}, t_k)$ and $\Phi_{rv}(t_{k-s}, t_k)$ are submatrices of the state transition matrix $\Phi(t_{k-s}, t_k)$ from epoch $t_k$ to $t_{k-s}$. The measurement Jacobian matrix is finally given by

$$H_k^{EDGG} = \begin{bmatrix} H_{r,k}^{EDGG} & H_{v,k}^{EDGG} \end{bmatrix} \tag{30}$$

It is noted that the accuracy requirement for the Jacobian matrix is not stringent and only the Earth's gravitation up to degree 2 and order 0 is involved for computations of $\Phi$ and $T_e$.

*3.2.2. Starlight refraction angle*

The measurement model of starlight refraction angle is given as follows

$$z_k^{SRA} \equiv \begin{bmatrix} R_{1,k} \\ R_{2,k} \\ \vdots \\ R_{N_k,k} \end{bmatrix} = \begin{bmatrix} \eta_1(r_k) + \xi_{1,k} \\ \eta_2(r_k) + \xi_{2,k} \\ \vdots \\ \eta_{N_k}(r_k) + \xi_{N_k,k} \end{bmatrix} = h^{SRA}(x_k, t_k) + \xi_k \tag{31}$$

where $N_k$ is the number of observed refracted stars at $t_k$, $R_{j,k}$, $j=1,2,...,N_k$ is the measured refraction angle for the $j$th star at $t_k$, $\eta_j(\cdot)$ represents the implicit function of Eq. (12), $\xi_{j,k}$ represents the observation noise and $h^{SRA}(\cdot)$ represents the SRA measurement function. $\xi_{j,k}$ is assumed to be white and Gaussian and independent of the refracted star. The covariance matrix of the noise vector $\boldsymbol{\xi}_k = \begin{bmatrix} \xi_{1,k} & \xi_{2,k} & \cdots & \xi_{N_k,k} \end{bmatrix}^T$ is denoted by $\boldsymbol{R}_k^{SRA}$, which is a $N_k \times N_k$ diagonal matrix.

The partial derivatives of $R$ with respect to $\boldsymbol{r}$ can be obtained by differentiating on both sides of Eq. (12)

$$\frac{\partial R}{\partial \boldsymbol{r}^T} = \frac{1}{1000} \cdot \frac{\dfrac{\boldsymbol{r} - (\boldsymbol{r} \cdot \boldsymbol{u}_s)\boldsymbol{u}_s}{\sqrt{r^2 - u^2}} + \dfrac{\partial u}{\partial \boldsymbol{r}^T} \tan R}{67.86 R^{-0.0195} - 6.44 R^{-1} - u(\cos R)^{-2}} \tag{32}$$

where the factor 1/1000 accounts for the unit of kilometer used in Eq. (12). The partial derivative matrix of $\boldsymbol{z}_k^{SRA}$ with respect to $\boldsymbol{r}_k$ is

$$\boldsymbol{H}_{r,k}^{SRA} = \begin{bmatrix} \dfrac{\partial R_{1,k}}{\partial \boldsymbol{r}_k} & \dfrac{\partial R_{2,k}}{\partial \boldsymbol{r}_k} & \cdots & \dfrac{\partial R_{N_k,k}}{\partial \boldsymbol{r}_k} \end{bmatrix}^T \tag{33}$$

The measurement Jacobian matrix is finally obtained as follows

$$\boldsymbol{H}_k^{SRA} = \begin{bmatrix} \boldsymbol{H}_{r,k}^{SRA} & \boldsymbol{0} \end{bmatrix} \tag{34}$$

The partial derivatives with respect to the velocity vector are all zero since that the velocity vector does not appear in the observation equation.

*3.3. Information fusion filter*

The information fusion of EDGG and SRA data is implemented using a sequential filter mechanism, as depicted in Fig. 3. The state vector and covariance are predicted using the orbital dynamic model from last epoch to current epoch. The EDGG measurements are first employed to correct the state prediction via the EKF. If there are any observed refracted stars at current epoch, the SRA measurements are subsequently employed to further correct the state estimates via the EKF. Compared to the commonly used federated filter,[24] the sequential information fusion filter can better handle the frequent data outages of SRA measurements. In addition, the sequential mechanism could be easily adapted for asynchronous measurements of the two subsystems.

Let $\hat{\boldsymbol{x}}_{k-1|k-1}$ and $\hat{\boldsymbol{P}}_{k-1|k-1}$ denote the estimated state vector and covariance at epoch $t_{k-1}$. The state prediction is given by

$$\hat{\boldsymbol{x}}_{k|k-1} = \boldsymbol{\varphi}\left( \hat{\boldsymbol{x}}_{k-1|k-1}, t_k, t_{k-1} \right) \tag{35}$$

$$\hat{\boldsymbol{P}}_{k|k-1} = \boldsymbol{\Phi}(t_k, t_{k-1}) \hat{\boldsymbol{P}}_{k-1|k-1} \boldsymbol{\Phi}(t_k, t_{k-1})^T + \boldsymbol{Q}_{k-1} \tag{36}$$

where $\hat{x}_{k|k-1}$ and $\hat{P}_{k|k-1}$ are the predicted state vector and covariance at epoch $t_k$. The first state correction from EDGG measurements is implemented as follows

$$\hat{x}_{k|k}^{EDGG} = \hat{x}_{k|k-1} + K_k^{EDGG}\left[z_k^{EDGG} - h^{EDGG}\left(\hat{x}_{k|k-1}, t_k\right)\right] \tag{37}$$

$$\hat{P}_{k|k}^{EDGG} = \left(I - K_k^{EDGG} H_k^{EDGG}\right)\hat{P}_{k|k-1}\left(I - K_k^{EDGG} H_k^{EDGG}\right)^T - K_k^{EDGG} R^{EDGG}\left(K_k^{EDGG}\right)^T \tag{38}$$

$K_k^{EDGG}$ is the EDGG gain and is given by

$$K_k^{EDGG} = \hat{P}_{k|k-1}\left(H_k^{EDGG}\right)^T\left[H_k^{EDGG}\hat{P}_{k|k-1}\left(H_k^{EDGG}\right)^T + R^{EDGG}\right]^{-1} \tag{39}$$

The further state correction from SRA measurements is

$$\hat{x}_{k|k}^{IF} = \hat{x}_{k|k}^{EDGG} + K_k^{SRA}\left[z_k^{SRA} - h^{SRA}\left(\hat{x}_{k|k}^{EDGG}, t_k\right)\right] \tag{40}$$

$$\hat{P}_{k|k}^{IF} = \left(I - K_k^{SRA} H_k^{SRA}\right)\hat{P}_{k|k}^{EDGG}\left(I - K_k^{SRA} H_k^{SRA}\right)^T - K_k^{SRA} R^{SRA}\left(K_k^{SRA}\right)^T \tag{41}$$

with

$$K_k^{SRA} = \hat{P}_{k|k}^{EDGG}\left(H_k^{SRA}\right)^T\left[H_k^{SRA}\hat{P}_{k|k}^{EDGG}\left(H_k^{SRA}\right)^T + R^{SRA}\right]^{-1} \tag{42}$$

$\hat{x}_{k|k}^{IF}$ and $\hat{P}_{k|k}^{IF}$ are the state vector and covariance at epoch $t_k$ after information fusion.

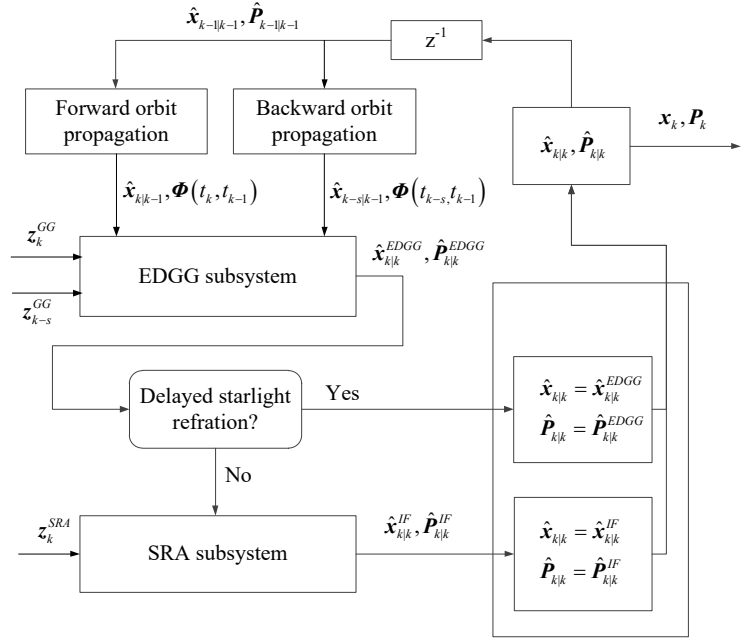

**Fig. 3.** Block diagram of the information fusion filter based on sequential processing mechanism.

## 4. Simulation results and analysis

In this section, the performance of integrated navigation using epoch-differenced gravity gradients and starlight refraction is tested with simulated data. Several impact factors such as the differencing interval and orbital height are also analyzed.

*4.1. Simulation conditions*

The simulation covers an 18-hour data arc starting from 5 December 2015, 12:00:00.0 (UTC Time). The truth orbit trajectory is simulated using a high-precision numerical orbit simulator, in which a 120 × 120 subset of the EGM2008 model for Earth's non-spherical gravitational attraction, the NRLMSISE-00 model for atmospheric density, the Jet Propulsion Laboratory (JPL) DE405 ephemeris for lunar and solar positions, and the Adams-Bashforth-Moulton method for numerical integration are used. The LEO satellite under consideration has an orbital height of 300 km. The initial osculating orbital elements are as follows: semi-major axis $a$ = 6678.14 km, eccentricity $e$ = 0, inclination $i$ = 60˚, right ascension of ascending node $\Omega$ = 120˚, the argument of perigee $\omega$ = 0˚, and the mean anomaly $M$ = 80˚.

The truth gravity gradients are generated using a 300 × 300 subset of the EGM2008 model. The GRF frame is assumed to be always aligned with the satellite Local Vertical Local Horizontal (LVLH) reference frame, where the X-axis is outward along the radial (local vertical), Y-axis is perpendicular to X in the orbit plane in the direction of motion (local horizontal), and Z-axis is along the orbit normal. Significant biases having a low drift rate of 0.01 E/h, noises near the orbit frequency with a magnitude of 0.1 E, as well as white noises with a standard deviation of 0.1 E are added to the truth gravity gradients in order to simulate noisy measurements. The precision of both the attitude star sensor and the refraction star sensor is assumed to be 1 arcsec, with a detectable stellar magnitude of 6.0 and a FOV of 10˚ × 10˚. According to Chen et al.[8], at the orbital altitude of 300 km, the GGT observation error caused by an attitude error of 1 arcsec is about 0.013 E, which is one order of magnitude lower than the noise level of the simulated gravity gradient measurements. Thus, the effect of attitude error on the EDGG subsystem can be neglected. The installation angle of the refraction star sensor is calculated according to Eq. (14). The Tycho-2 catalogue[25] is used as the reference catalogue. As stated earlier, the starlight refraction model error is about 0.2 arcsec and is not considered. The measurements are simulated with a data-sampling period of 30 s.

The main filter parameters are set as follows. The initial position and velocity errors are set to [10 km, 10 km, 10 km, 10 m/s, 10 m/s, 10 m/s] and the diagonal elements of the initial covariance of the state vector are set to [(10 km)$^2$, (10 km)$^2$, (10 km)$^2$, (10 m/s)$^2$, (10 m/s)$^2$, (10 m/s)$^2$]. The diagonal elements of the covariance matrix $\boldsymbol{R}^{GG}$ is set to [(0.10 E)$^2$, (0.10 E)$^2$, (0.10 E)$^2$, (0.10 E)$^2$, (0.10 E)$^2$, (0.10 E)$^2$]. The standard deviation of the process noise $\boldsymbol{w}_t$ is set to $5 \times 10^{-4}$ m/s$^2$ and the standard deviation of $\xi_{j,k}$ is set to 1 arcsec. The differencing interval is set to 5.

*4.2. Integration performance analysis*

This section compares the performances of orbit determination using solely EDGG, solely SRA, and EDGG + SRA measurements. The time varying position and velocity estimation errors are shown in Fig. 4 and 5, respectively, and the RMS values of the steady-state estimation errors are listed in Table 1. The statistical three-dimensional (3D) error is the root of the sum of the squares of the radial, along-track and cross-track errors. Compared to the orbit determination results presented in Ref.[10], the epoch-differencing strategy achieves similar and even better navigation accuracy. Especially, the radial and cross-track position errors are improved. This indicates the effectiveness of the epoch-differencing strategy to deal with significant biases and low-frequency noises. Among the three orbit determination methods, the integrated navigation filter converges with the fastest speed and achieves the best accuracy. The 3D root mean square (RMS) position and velocity errors for the integrated navigation are 69.175 m and 0.0771 m/s, respectively. Compared to solely EDGG orbit determination, the information fusion significantly reduces the along-track position error from 886.41 m to 66.145 m and the radial velocity error from 1.0237 m/s to 0.0735 m/s. It should be noted that the effective accuracy improvements of these two components are consequences of simultaneous bias and noise reduction. Compared to solely SRA orbit determination, the information fusion improves the radial and cross-track position accuracies as well as the along-track and cross-track velocity accuracies. Especially, the cross-track position and velocity errors are reduced from 159.58 m and 0.1825 m/s to 15.301 m and 0.0185 m/s, respectively. Overall, the information fusion filter makes the most of both EDGG and SRA measurements and handles the weights of these two kinds of observations properly. As a result, the integrated orbit determination strategy excels the method based on either gravity gradient or starlight refraction solely.

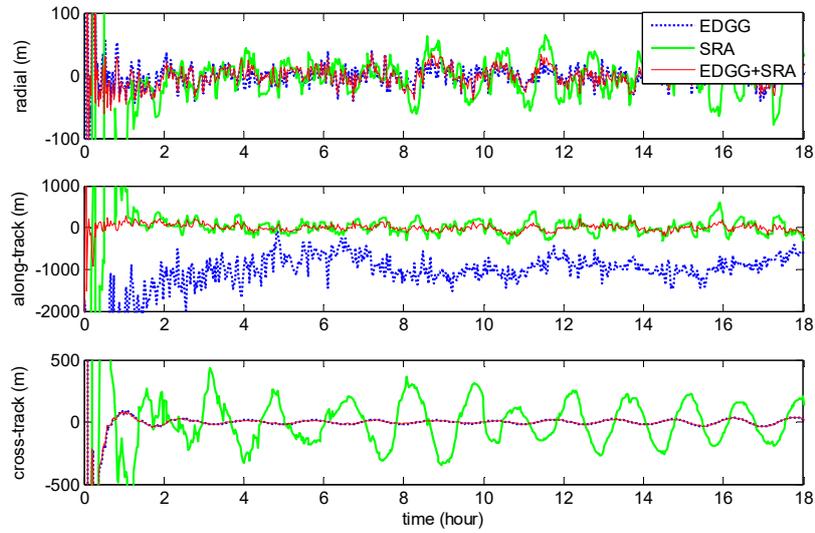

**Fig. 4.** Time varying position estimation errors for the three orbit determination strategies.

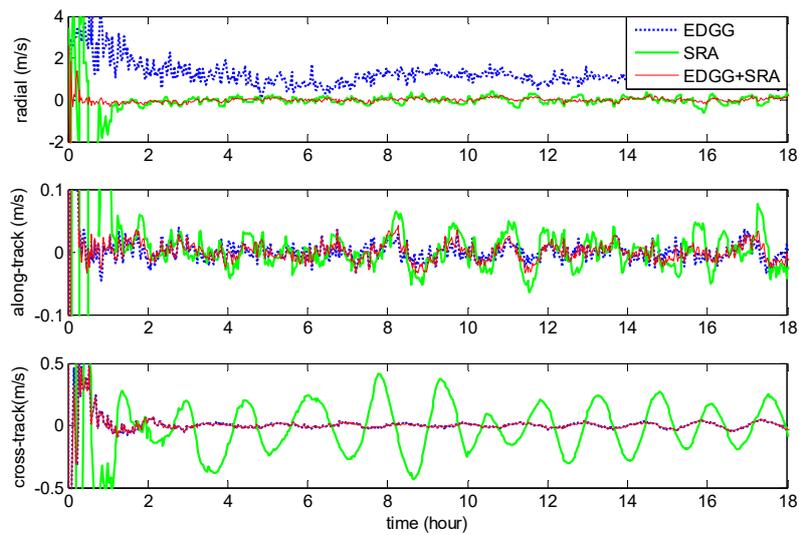

**Fig. 5.** Time varying velocity estimation errors for the three orbit determination strategies.

**Table 1**. RMS values of steady-state position and velocity errors for the three orbit determination strategies

| Observation | Position error, m | | | | Velocity error, m/s | | | |
|---|---|---|---|---|---|---|---|---|
| | Radial | Along-track | Cross-track | 3D | Radial | Along-track | Cross-track | 3D |
| EDGG | 13.207 | 886.41 | 16.180 | 886.66 | 1.0237 | 0.0120 | 0.0187 | 1.0239 |
| SRA | 26.430 | 153.01 | 159.58 | 222.66 | 0.1604 | 0.0244 | 0.1825 | 0.2442 |
| EDGG+SRA | 13.136 | 66.145 | 15.301 | 69.175 | 0.0735 | 0.0129 | 0.0185 | 0.0771 |

*4.3. The influence factors analysis*

*4.3.1 Effects of differencing intervals*

Orbit determination using EDGG + SRA measurements with different differencing intervals ($s$ = 1, 2, 5, 10, and 20) is implemented in order to explore the effects of epoch-differencing intervals. The RMS values of the steady-state position and velocity estimation errors are summarized in Table 2. It is seen that the orbit determination accuracy improves with the increase of differencing interval. The reason lies in the fact that increasing the differencing interval augments the amount of the remaining effective information of gravity gradients about the satellite orbit after the differencing operation. As seen in Table 2, the 3D position and velocity errors for $s$ = 20 are about half of those for $s$ = 1. However, a larger differencing interval requires a longer computation time of the modeled EDGG measurements. Thus the value of $s$ should be compromised in navigation algorithm design.

**Table 2**. RMS values of steady-state position and velocity errors with different differencing intervals

| Differencing interval | Position error, m | | | | Velocity error, m/s | | | |
|---|---|---|---|---|---|---|---|---|
| | Radial | Along-track | Cross-track | 3D | Radial | Along-track | Cross-track | 3D |
| 1 | 15.442 | 85.857 | 27.165 | 91.367 | 0.0914 | 0.0159 | 0.0328 | 0.0984 |
| 2 | 14.670 | 81.334 | 15.825 | 84.148 | 0.0871 | 0.0152 | 0.0196 | 0.0905 |
| 5 | 13.136 | 66.145 | 15.301 | 69.175 | 0.0735 | 0.0129 | 0.0185 | 0.0771 |
| 10 | 14.168 | 58.136 | 12.497 | 61.128 | 0.0691 | 0.0129 | 0.0149 | 0.0719 |
| 20 | 12.524 | 42.940 | 13.104 | 46.380 | 0.0532 | 0.0121 | 0.0150 | 0.0566 |

*4.3.2 Effects of orbital height*

The spherical shape of the Earth's gravitational field indicates that the accuracy of the gravity gradient based orbit determination is highly dependent of the orbital height and the eccentricity but has little relationship with other orbital elements, such as orbital inclination, argument of perigee, etc. In fact, the eccentricity influences orbit determination accuracy due to the fact that it makes the orbital height varies with time. In addition, the number of visible refracted stars per orbit period also varies with orbital height. Thus, only the effects of orbital height is analyzed. Besides the 300 km height, another four cases in which the orbital heights are 600 km, 1000 km, 1500 km, and 2000 km are also simulated to analyze the effects of orbital heights on integration navigation accuracy.

The RMS values of steady-state position and velocity estimation errors are shown in Table 3. It is seen that the orbit determination accuracy decreases with increasing the orbital height. To be specific, the 3D position error of the 2000 km case is 166.69 m which is about twice that of the 300 km case. When the orbital height further increases to geosynchronous orbits, the orbit determination error could be unfavorable or unacceptable. This phenomenon can be explained as follows. Firstly, the sensitivity of gravity gradient signals with respect to position decreases with height. The sensitivity factor of GGT has been defined in Ref.[8] and can be approximately given by $3GM/r^4$. Secondly, the number of visible refracted stars is negatively related to the orbital height and the orbit estimation accuracy decreases as the visible stars get less. The values of GGT sensitivity factors as well as the average number of visible stars per orbit are also given in Table 3. The values of these two accuracy factors decrease from $6.013 \times 10^{-4}$ E/m and 182 to $2.427 \times 10^{-4}$ E/m and 88 when the orbital height increases from 300 km to 2000 km.

Table 3. RMS values of steady-state 3D position and velocity errors as well as accuracy factors with different orbital heights

| Orbital height, km | Position error, m | Velocity error, m/s | GGT sensitivity factor, E/m | Average number of visible stars per orbit |
|---|---|---|---|---|
| 300 | 69.175 | 0.0771 | $6.013 \times 10^{-4}$ | 182 |
| 600 | 75.669 | 0.0772 | $5.044 \times 10^{-4}$ | 130 |
| 1000 | 95.201 | 0.0804 | $4.036 \times 10^{-4}$ | 109 |
| 1500 | 130.25 | 0.1073 | $3.105 \times 10^{-4}$ | 94 |
| 2000 | 166.69 | 0.1184 | $2.427 \times 10^{-4}$ | 88 |

## 5. Semi-simulation with real GOCE GGT data

The performance of the integrated navigation filter has also been tested with real GGT data from GOCE combined with simulated SRA measurements. The GOCE was launched on 17 March 2009 into a target sun-synchronous near-circular orbit with a height of 278.65 km. The spaceborne gravity gradiometer was designed to measure gravity gradients. A stable and quiet measuring environment was ensured by an electric propulsion engine compensating for non-gravitational forces. Three star sensors and two dual-frequency geodetic GPS receivers were carried to provide high-precision attitude and orbit information.[26]

An 18-hour data arc beginning from 8 September 2013, 00:00:00.0 (GPS Time) is used for the test. The data on this day are reported to have good quality. The measurements are resampled at an interval of 30 s. Data analysis in Ref.[9] showed that in the measurement bandwidth the white noise density levels are on the order of 10 mE/√Hz for the $\Gamma_{xx}^{(g)}$, $\Gamma_{yy}^{(g)}$, $\Gamma_{zz}^{(g)}$, and $\Gamma_{xz}^{(g)}$ components, whereas for the $\Gamma_{xy}^{(g)}$ and $\Gamma_{yz}^{(g)}$ components, the white noise density levels are 350 and 500 mE/√Hz, respectively. A refraction star sensor is assumed onboard GOCE to provide SRA measurements. The same simulation conditions as those stated in Section 4.1 are used for starlight refraction.

The initial position and velocity errors are set to 10 km and 10 m/s. The standard deviation of the process noise $w_t$ is set to $5 \times 10^{-4}$ m/s². The diagonal elements of the covariance matrix $\boldsymbol{R}^{GG}$ is set to [(0.01 E)², (0.01 E)², (0.01 E)², (0.35 E)², (0.01 E)², (0.50 E)²]. The standard deviation of $\xi_{j,k}$ is set to 1 arcsec.

The GPS-derived orbits serve as true values for accuracy analysis. The time-varying position and velocity estimation errors are shown in Fig. 6 and 7, respectively. The RMS values of steady-state position and velocity errors are summarized in Table 4. Similar to the simulation results, the integrated navigation filter converges the fastest and achieves the best accuracy for the GOCE semi-simulation. The 3D RMS position and velocity errors for the integrated navigation are 98.925 m and 0.1116 m/s, respectively. Compared to solely EDGG orbit determination, the integration reduces the along-track position error from 591.34 m to 88.538 m and the radial velocity error from 0.6997 m/s to 0.0961 m/s. Compared to solely SRA orbit determination, the integration improves the radial and cross-track position accuracies as well as the along-track and cross-track velocity accuracies.

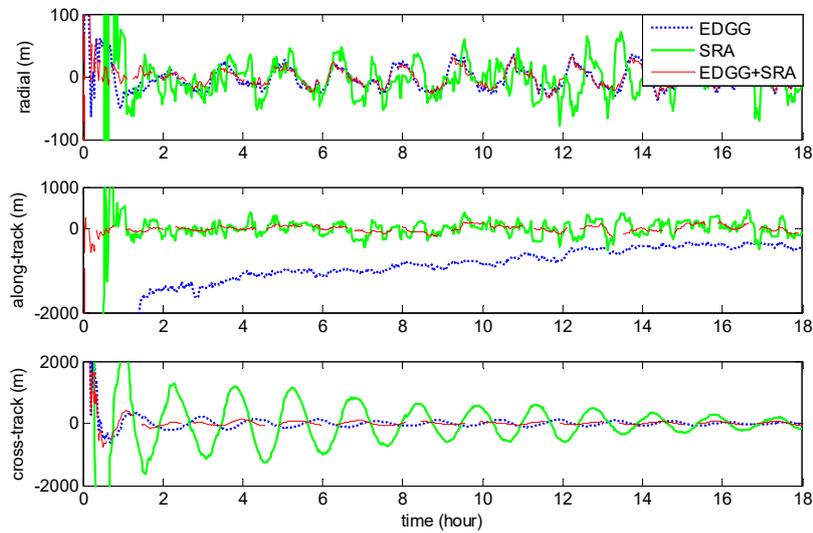

**Fig. 6.** Time varying position estimation errors for the GOCE semi-simulation case.

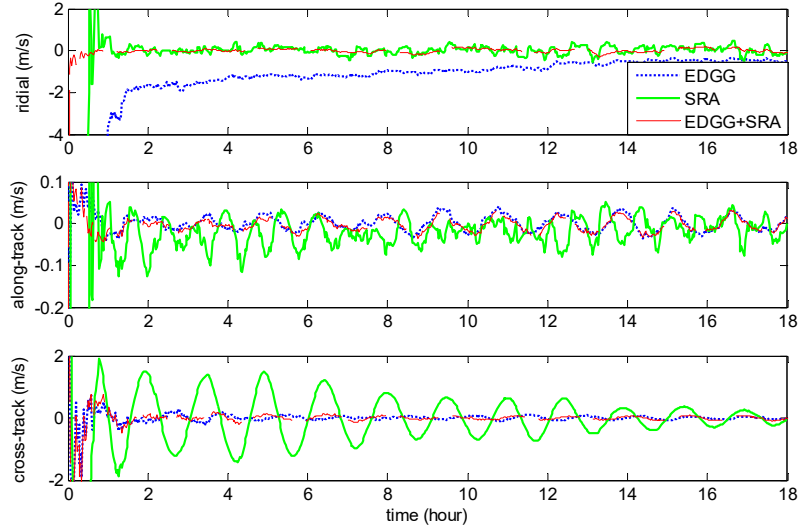

**Fig. 7.** Time varying velocity estimation errors for the GOCE semi-simulation case.

**Table 4**. RMS values of steady-state position and velocity errors for the GOCE semi-simulation case

| Observation | Position error, m | | | | Velocity error, m/s | | | |
|---|---|---|---|---|---|---|---|---|
| | Radial | Along-track | Cross-track | 3D | Radial | Along-track | Cross-track | 3D |
| EDGG | 17.377 | 591.34 | 58.684 | 594.50 | 0.6979 | 0.0187 | 0.0463 | 0.6997 |
| SRA | 24.898 | 159.63 | 132.81 | 209.15 | 0.1884 | 0.0284 | 0.2040 | 0.2790 |
| EDGG+SRA | 16.153 | 88.538 | 41.065 | 98.925 | 0.0961 | 0.0167 | 0.0543 | 0.1116 |

## 6. Conclusions

The performance of integration of gravity gradiometry and starlight refraction for LEO autonomous orbit determination has been demonstrated in this study. The integration is implemented by an information fusion filter based on a sequential EKF mechanism which better addresses asynchronous measurements of multiple sensors. The gravity gradients are time differenced to eliminate slowly varying measurement biases and noises near the orbit revolution frequency. The refraction angle is directly used as measurement and its Jacobian matrix has been derived from an implicit observation equation. The method significantly improves the along-track position accuracy compared to that using solely gravity gradients and improves the radial and cross-track position accuracies compared to that using solely starlight refraction.

The proposed orbit determination method is immune to signal blockage and spoofing encountered in GPS navigation. Compared to other navigation approaches such as the magnetic field based navigation and the X-ray pulsar based navigation, which have typical position errors of a few kilometers, better accuracy of tens of meters have been achieved.